\documentclass[12pt,letterpaper]{article}
\usepackage{amsmath,amssymb,pgf,pgfarrows,pgfnodes,float,appendix, hyperref}
\usepackage{graphicx}
\usepackage{subfigure}
\usepackage[margin=0.9in]{geometry}
\usepackage{tikz}
\usepackage{cite}

\newcommand{\be}{\begin{equation}}
\newcommand{\ee}{\end{equation}}
\newcommand{\bea}{\begin{eqnarray}}
\newcommand{\eea}{\end{eqnarray}}

\title{{\rm\footnotesize \qquad \qquad \qquad \qquad \qquad \ \qquad \qquad \qquad \ \ \ \ \ \                 RUNHETC-2025-36}\vskip.5in   JT de Sitter Gravity as a Model of Coleman-de Luccia Tunneling}
\author{Sidan A\\
Department of Physics and NHETC\\
Rutgers University, Piscataway, NJ 08854\\
E-mail: \href{mailto:sa1975@physics.rutgers.edu}{sidan.aa@rutgers.edu}
\\
\\
Tom Banks\\
Department of Physics and NHETC\\
Rutgers University, Piscataway, NJ 08854\\
E-mail: \href{mailto:tibanks@ucsc.edu}{tibanks@ucsc.edu}
\\
\\
}
\date{}
\begin{document}
\maketitle

\begin{abstract} We interpret the Euclidean solution of JT de Sitter gravity as the Coleman-de Luccia instanton for the decay of the low-entropy horizon of its static patch solution into either a Big Crunch or an infinite-entropy Lorentzian de Sitter cosmology. As in previous work by one of the authors, the Big Crunch is interpreted as bounding the entropy of the static state. The principle of detailed balance then guarantees it will transition back to a higher entropy causal diamond in the expanding cosmology. We then construct a family of explicit quantum mechanical models and appropriate metastable states, with transition probabilities well approximated by the semi-classical calculations in JT de Sitter gravity.\end{abstract}
\maketitle

\newpage
\section{Introduction} 

Jackiw-Teitelboim gravity \cite{Jackiw:1984je,Teitelboim:1983ux} with negative cosmological constant (c.c.) has been an extremely useful toy model for exploring general ideas about quantum gravity. There have also been a number of papers written about the version of the model with positive c.c. \cite{Maldacena:2019cbz, Cotler:2019nbi, Cotler:2019dcj, Moitra:2022glw, Susskind:2022dfz, Rahman:2022jsf, Susskind:2022bia, Cotler:2023eza, Susskind:2023hnj, Susskind:2023rxm, Nanda:2023wne, Rahman:2023pgt, Cotler:2024xzz, Rahman:2024vyg, Rahman:2024iiu, Alonso-Monsalve:2024oii, Held:2024rmg, Dey:2025osp, Sekino:2025bsc, Okuyama:2025hsd}.  This is another one.  The physical interpretation we will propose is radically different from that in any previous work.  We have however found the presentation of the classical Lorentzian solutions of the theory in \cite{Maldacena:2019cbz} extremely useful.  We will argue that two of them can both be viewed as analytic continuations of a single Euclidean solution.  That solution has the form of a two sphere, with a dilaton field that varies like $\sin (\theta )$, where $\theta$ is the polar angle.  This has the form of an instanton for the decay of the static patch of dS space, with the peculiar feature that the initial and final states have the same value of the c.c., {\it but different entropies}.   In making this statement we are making the usual identification of the field that multiplies $\sqrt{-g}R$ in the $1 + 1$ dimensional gravitational action as the entropy \cite{cghs}.  

Indeed we find that three analytic continuations of the Euclidean solution to Lorentzian signature can be interpreted as
\begin{itemize}
\item A static patch of dS space with a spatially varying entropy field.
\item A decay of the static patch to a finite entropy Big Crunch.
\item A decay of the static patch to an expanding Big Bang dS cosmology with time dependent unbounded entropy.  
\end{itemize}

Our proposal is thus that JTdS is a hydrodynamic description of a quantum system with a metastable state with these properties.  In the third section of the paper, we exhibit an explicit class of quantum systems that behave in the manner suggested by JTdS gravity.  The Poincar\'e solution of JTdS exhibited in \cite{Maldacena:2019cbz} uses the flat slicing of dS space and has a dilaton field that varies as
\begin{equation} 
    S_{\text{Poincar\'e}} = S_0 + \frac{S_1}{\eta} . 
\end{equation}
The conformal time $\eta$ is related to the space and time coordinates of the Milne cosmology that we will exhibit below by
\begin{equation} 
    \frac{1}{\eta} = \sinh (\tau ) \cosh (\chi )-\cosh (\tau ) . 
\end{equation}
The dilaton field blows up at $\chi = \pm \infty$ and the solution is thus eliminated by any reasonable boundary conditions on the other solutions, such as those that follow naturally from analytic continuation. Since the Poincar\'e solution is just a half-coordinate patch of the global solution, related by a coordinate transformation, the global solution is also ruled out by the boundary conditions. As a consequence, all of the Lorentzian solutions are invariant under translations in $\chi$ and we are free to view this as a gauge symmetry of the problem even though the range of the variable is non-compact.  This will no longer be true if we couple the model to systems with true field theoretic degrees of freedom.  In the following sections, we describe formalisms for treating both the possibility of translation gauge symmetry and one in which there might be excitations that are localized in $\chi$. 

Throughout this paper we will be interpreting the $S$ function as describing the entropy in some causal diamond.  This is in accord with what we call the {\it Covariant Entropy Principle} (CEP) in higher dimensions, and the fact that $S$ arises from the volume of a compact $d - 2$ fold when a two dimensional theory is obtained from a $d$ dimensional one by compactification.  The CEP states that the entropy in a causal diamond is $\frac{A_{\diamond}}{4G_N}$, where $A_{\diamond}$ is the maximal $d - 2$ volume on the diamond boundary.  This principle can be viewed as lying behind Jacobson's \cite{ted95} derivation of Einstein's equations from hydrodynamics. For black holes it has been ``derived" by Carlip \cite{carlip} and Solodukhin \cite{solo}, and \cite{BZ} extended that derivation to arbitrary causal diamonds.  For causal diamonds in maximally symmetric space-times the CEP has been derived by Euclidean path integral calculations \cite{BDF}.  At any rate, it is the basic assumption behind our work.

We would like to point out that Maldacena et. al. \cite{Ivo:2024ill} have recently considered a proposal for using Coleman-de Luccia (CdL) instantons to compute contributions to the density matrix of subregions of the universe. Their proposal has some similarities to our explicit model but we do not understand the connection. In addition, the vacuum transition amplitude of two-dimensional CdL was computed by Pasquarella and Quevedo \cite{Pasquarella:2022ibb}, where the principle of detailed balance is also applied to address the entropy of the dS static patch. In their paper, the transition happens when two universes have different cosmological constants. Their interpretation differs from ours, in our case, it is not the c.c. but the entropy function that matters in 1+1 dimensions. 

\section{The Instanton and Its Continuations}

The Euclidean solution of JTdS gravity is
\begin{equation} 
    ds^2 = d\theta^2 + \sin^2 (\theta ) d\phi^2  , 
\end{equation}
\begin{equation} 
    S = S_0 + S_1 \cos (\theta ) . 
\end{equation}
This has the form of a Coleman-de Luccia (CdL) instanton because $S^{\prime} (0) = S^{\prime} (\pi) = 0 $.  It is peculiar in that the c.c. does not change between the two poles of the sphere, but crucially {\it the entropy does change}.   Thus the interpretation of CdL transitions in terms of thermal physics advocated in \cite{Banks:2002nm, Banks:2005ru, Aguirre:2006ap, Brown:2007sd, Banks:2021wqu} can be applied.  Here we are assuming the interpretation of the $S$ field as entropy \cite{Maldacena:2019cbz, cghs}.  

The metastable dS space that is decaying is revealed by the analytic continuation $\phi \rightarrow i t $ and gives the static patch Lorentzian solution.
\begin{equation} 
    ds^2 = d\theta^2 - \sin^2 (\theta ) dt^2  ,
\end{equation}
\begin{equation} 
    S = S_0 + S_1 \cos (\theta ) . 
\end{equation}
Positivity of entropy implies $S_0 \pm S_1 \geq 0$, where $0$ indicates a singularity. We assume that, it is at the left pole $\theta = 0$ that the static patch transitions into the expanding dS cosmology, and at the right pole $\theta = \pi$ that it transitions into a Big Crunch. 

The two states into which the static patch decays are given, as usual, by the analytic continuation $\theta \rightarrow i \tau$ and $ \theta - \pi \rightarrow i \tau$. We also need to analytically continue $\phi \rightarrow i\chi$ to obtain the metric in Lorentzian signature. These lead to 
\begin{equation} 
    ds_{\text{Milne}}^2 = - d\tau^2 + \sinh^2 (\tau) d\chi^2 . 
\end{equation}
\begin{equation}  
    S_{\text{Milne}} = S_0 + S_1 \cosh (\tau ) , \ \ \tau \geq 0 , 
\end{equation}
and 
\begin{equation} 
    ds_{\text{Crunch}}^2 = - d\tau^2 + \sinh^2 (\tau ) d\chi^2 , 
\end{equation}
\begin{equation} 
    S_{\text{Crunch}} = S_0 - S_1 \cosh (\tau ), \ \ \ \tau \leq 0 . 
\end{equation}
This tells us that the spatial coordinate $\chi \in \{-\infty, \infty\}$ in both the Milne and the Big Crunch solutions. In order for the Milne solution to be non-singular for $\tau > 0$ we must have $S_1 > 0$, in which case the Crunch solution has a singularity at finite proper time. This of course implies that $S_0$ is also positive and $S_0 \geq S_1$. 

If we recall the prescription of \cite{Banks:2002nm, Banks:2005ru, Aguirre:2006ap, Brown:2007sd, Banks:2021wqu}, the transition probabilities are computed by adding to the negative instanton action the entropies of the two final states, i.e. $S_{\text{Milne}}$ and $S_{\text{Crunch}}$, so that the relative probabilities satisfy the principle of detailed balance.  The Big Crunch is interpreted as the breakdown\footnote{When the entropy becomes so low that the semiclassical approximation is no longer valid.} of the hydrodynamic description in a low entropy state, just before the transition back to the higher entropy system.  What is novel about the situation at hand is that one of the Lorentzian systems does not have a time independent Hamiltonian.  

Time dependent Hamiltonians in quantum gravity are the analog of half sided modular inclusions in quantum field theory.  If one believes in the idea of a causal diamond as a subsystem and that the covariant entropy principle has any precise quantum mechanical meaning, then operator algebras associated with diamond subsystems cannot be Type III. There are indications \cite{yngvason, Yngvason:2014oia} that systems with time independent Hamiltonians can only have sharp light cones if the spectrum of the Hamiltonian admits infinite energy states. This means that the only way that diamonds with finite dimensional Hilbert spaces can truly be subsystems\footnote{Take a lattice field theory with a time independent Hamiltonian as an example, a finite region of space in such a theory is a subsystem only for an infinitesimal period of time.  Any finite time translation operator connects it to regions that are arbitrarily far away, albeit with exponentially small amplitude.} is if the Hamiltonian is time dependent. We note that the time-band sub-algebras of Liu and Leutheusser \cite{LL} exist only in the strict $N = \infty$ limit, and are Type III.  

With time dependent unitary evolution it's easy to construct exactly causal dynamics \cite{HST04,HST11,HST18,HST2502,HST2505}. These flows can be thought of as quantum gravity generalizations of the one sided modular inclusions of algebraic quantum field theory.  If we think of diamond Hilbert spaces as growing in dimension with the area of the diamond's holographic screen, then the time dependent mapping is a unitary embedding of a small diamond Hilbert space into that of a larger diamond, the proper time from some common origin of whose future tip, along some particular geodesic, is larger by one Planck unit than the smaller diamond (see Fig. \ref{fig:nested} in Appendix). 
This embedding must be supplemented by a unitary transformation in the commutant of the larger diamond operator algebra, to give a unitary on the full Hilbert space of the model.  

This sort of time evolution has been used extensively in holographic space-time models for almost a quarter of a century \cite{HST04,HST11,HST18,HST2502,HST2505}.  At each time, one factors the Hilbert space into ${\cal H}_{in} \times {\cal H}_{out}$.  The time evolution operator is written as
\begin{equation} 
    U(t) = U_{in} (t) \otimes U_{out} (t) . 
\end{equation} 
$U_{in} (t)$ is the unitary embedding between consecutive nested diamond Hilbert spaces described above (the analog of one sided modular inclusion).  The authors of \cite{HST04,HST11,HST18,HST2502,HST2505} conjectured that $U_{out} (t)$ is determined by the {\it Quantum Principle of Relativity}:  One constructs independent time evolutions along all the geodesics in some background metric, and requires that the dynamics be such that one can consistently assign the a density matrix with the same entanglement spectrum to the maximal causal diamond in the overlap between any two diamonds along an arbitrary pair of geodesics. Further details can be found in \cite{HST04,HST11,HST18,HST2502,HST2505}.

Now imagine the system in the metastable state represented by the static patch solution.  Notice that unlike the conventional static coordinates on dS space, this coordinate patch has two horizons, of differing entropy.  This is analogous to the Schwarzschild-de Sitter black hole in $4$ or more dimensions.  That solution has perturbative instabilities indicated by the higher temperature of the black hole horizon.  It emits gravitons and disappears back to the stable equilibrium of empty dS space.  In JTdS there are no particle excitations to emit and the low entropy horizon decays by tunneling.  The system it decays into has a time dependent density matrix.  At early times its density matrix is a tensor product of a large number of copies of a density matrix of very low entropy\footnote{The full Hilbert space of JTdS gravity has infinite dimension, as indicated by the asymptotic value of $S(\tau)$ in the Milne solution.  When we refer to tensor products we are always imagining the causal diamond Hilbert space at some large time $T$, where $T/L_P$ is taken to infinity at the end of the calculation. We assume that number theoretic issues having to do with when we can factorize the dimension of the full Hilbert space into products of the dimensions of finite diamond Hilbert spaces, become irrelevant in the limit.}.  This is the usual assumption that there are no correlations between causally disconnected regions of the universe built into the initial conditions of cosmology.  Without such an assumption the issue of ``the Horizon Problem" would never arise.  

We now have to make a postulate about how we want to view the finite entropy system represented by the static patch, in the full Hilbert space.  From the global God's eye view of quantum field theory, in which one imagines there is a global ``vacuum" state of a time independent Hamiltonian, and localized excitations on a pre-ordained time slice, we might want to consider them as localized excitations, which only occur at particular positions. But in a quantum theory that is trying to incorporate a quantum version of general covariance, there is no way to define a particular position without explicit local excitations in one's model.  The JTdS action has no local field theoretic degrees of freedom and so, by itself, has no local excitations.  Instead, we view the Milne solution as telling us that the hydrodynamic state of the quantum system under study is translation invariant in the $\chi$ direction.  Then the interpretation of the static patch is obvious.  At any given time, when the causal diamond in the Milne evolution is finite, there is a finite probability for the Milne evolution to make a transition to the static state.  The Milne entropy starts at the maximum of the static patch entropy $S_0+S_1$.  The Lorentzian time evolution from the low entropy side of the instanton ends in a Big Crunch.  Crunches have been interpreted \cite{tbcrunch} as a signal of the breakdown of hydrodynamic Einstein equations in low entropy situations and an indicator that the low entropy plus instanton action is an estimate of (minus) the logarithm of the probability of transition from the low to high entropy state.  We interpret the instanton computations as the probabilities at zero Milne time that one factor in the tensor product makes a transition between the Milne universe and the low entropy state in the static dS ``black hole" solution. 

\section{Quantum Mechanical Models}

We now want to build the simplest quantum models consistent with the hydrodynamic information encoded in JT dS gravity.  We first build a system describing the expanding cosmology.  This system has an infinite number of time-like geodesics, related by a translation symmetry.  Since the singularity at $\tau = 0$ restricts us to 
a single coordinate patch, we believe we must consider the spatial universe to be non-compact, so that the question of whether the translation is a gauge symmetry is open.  The Poincar\'e and the global solutions of the equations of motion are not invariant under translations.  For the purposes of the present paper we will ignore these solutions. In this case, all geodesics in the Milne cosmology are equivalent for all time, so we can model the system by concentrating on a single nested family of causal diamonds. 

The time dependence of the entropy function tells us the expectation value of the modular Hamiltonian of the system at each proper time.  Since the entropy is finite, we can realize it in a space of fermionic creation and annihilation operators $\psi_i$. According to the solution to JTdS gravity, the entropy begins at $\tau = 0$ at $S_0 + S_1$, which is already a macroscopically large number.  

If we write the modular Hamiltonian as an expansion in even powers of fermion operators, then for large degrees of freedom $N$ it is well known that the quadratic term is close to the Hamiltonian of a free $1 + 1$ dimensional Dirac fermion \cite{kwetal}, if the modular Hamiltonian is chosen at random.  Furthermore, with the exception of the choice of sign of the quartic term, all small perturbations of this do not disturb the spectrum except near the ``UV cutoff", where it is sparse.  Thus, there is an open set in the space of modular Hamiltonians where we can model the Milne universe by a sequence of quadratic fermion Hamiltonians $\psi_i K_{ij} \psi_j$ with random large $N$ Hermitian kernels.  JTdS gravity has only this hydrodynamic information and no localized excitations, so there is little more one can say 
without proposing a more specific low energy model with matter fields.

To compute the tunneling amplitudes between the initial state of the Milne universe and the low entropy state of the static black hole, we invoke the principle of detailed balance.  
The squared transition matrix elements for the back and forward transitions are equal and the relative probabilities are determined by Fermi's Golden Rule.  Let $S_{if}$ denote the transition matrix element between some state in the low entropy ensemble of the static patch and some final state in the initial $\tau=0$ state ensemble for the Milne universe. Then Fermi's Golden Rule states that the transition rate is given by
\begin{equation} 
    R(i \rightarrow \text{Milne}) = \sum_f |S_{if}|^2 W_f , 
\end{equation} 
where $W_f$ is some weighting of the final states, which depends on ``the detector efficiency in the Milne universe".  If we choose the basis for the Milne states in which the $\tau=0$ state density matrix is diagonal then 
\begin{equation} 
    \rho_f = \sum_f \frac{W_f}{\sum_k W_k} | f \rangle \langle f | , 
\end{equation} is the initial state density matrix of the Milne universe.  

In a similar fashion the inverse transition rate is given by (using unitarity)
\begin{equation} 
    R(f \rightarrow \text{Static} ) = \sum_i |S_{if}|^2 w_i , 
\end{equation} 
where $w_i$ is some weighting of the states in the static patch, and
\begin{equation} 
    \rho_i = \sum_i \frac{w_i}{\sum_j w_j} | i \rangle \langle i | , 
\end{equation} 
is the density matrix of the low entropy state of the static patch. In writing these equations we've made the assumption that in the present context, the ``detector efficiency" simply means that states are weighted in the same ratio as they are in the density matrix.  

The ratio of the two rates is given by 
\begin{equation} 
    r = \frac{\text{tr} (w)}{\text{tr} (W)} , 
\end{equation} 
where we've used an obvious notation for the un-normalized weight matrices.  We now assume that the modular Hamiltonians of the two systems belong to the large N universality class described above, which are well approximated by operators quadratic in fermions.  In this case the traces are products of a large number of factors.  The low entropy constraint on the static patch simply means that most of those factors in the numerator are simply $ = 1$, while the corresponding denominator factors are $ 1 + e^{-\lambda_a}$.  Therefore, the ratio is
\begin{equation} 
    r = \prod_{n=1}^{N_> - N_<} (1 + e^{-\lambda_n})^{-1} \prod_{m=1}^{N_<} \frac{1 + e^{-\kappa_m}}{1 + e^{- \lambda_m}}, 
\end{equation}  
where $\kappa_m, \lambda_m$ are the eigenvalues of $w, W$, respectively. $N_>$ is the number of q-bits in the initial Milne universe and $N_{<}$ the number in the low entropy state of the static patch.  

The {\it principle of detailed balance} is usually stated for thermal states and the ratio of forward and inverse transition probabilities is the exponential of the difference in {\it free energies}, divided by the temperature .   In the case of CdL transitions between two dS states the expectation values of the energy divided by the temperature  are numbers of order $1$, while the entropies are large, so these are just entropy differences.  The transition we are studying here is more analogous to the decay of a Schwarzschild-dS black hole into the dS vacuum, though in the JTdS model the dS vacuum is just the initial condition for the Lorentzian evolution to an infinite entropy state.  It's worth recalling that in $1 + 1$ dimensional gravity, Hawking temperatures are microscopic, entropy independent parameters, while black hole entropy and energy are proportional to each other.  The JTdS model does not have an asymptotic time-like or null boundary on which one could define energy rigorously, but these analogies make it entirely reasonable to interpret our detailed balance calculation for a large class of states in terms of the parameters of the JTdS instanton.   The actual instanton action would then be used to fix the parameter $S$ defined above, it incorporates all of the complications of the actual transition process, including the possibility that the $K_{ij}$ matrices for the initial Milne universe and the low energy state of the static patch are not diagonal in the same basis, in a single number.  JTdS gravity is thus a {\it very} coarse-grained hydrodynamic description, compatible with an infinite number of microscopic systems with different dynamics.  

In particular, we can choose {\it any} time dependent evolution with the property that at any proper time $t$ it is a unitary embedding of a Hilbert space of dimension $e^{S_0 + S_1 \cosh (t/R)}$ into one of dimension  $e^{S_0 + S_1 \cosh ([t + L_P]/R)}$  tensored with a unitary on some much larger space.  Given the parameters $S_{0,1}$ and dS radius $R$ defining the classical JTdS action we define the time step $L_P$ so that this shift is always an integer.  That is
\begin{equation} 
    \ln \left(1 + \frac{K_t}{N}\right) = S_1 \left[\cosh \left(\frac{t + L_P}{R}\right) - \cosh \left(\frac{t}{R}\right) \right] \approx \frac{S_1L_P}{R} \sinh \left(\frac{t}{R}\right), 
\end{equation} 
where $N$ and $N + K_t$ are the dimensions of the Hilbert spaces at times $t$ and $t + L_P$ respectively.  This is solved by
\begin{equation} 
    \frac{K_t}{N} = \frac{S_1L_P}{R} \sinh \left(\frac{t}{R}\right) \ll 1 , 
\end{equation} 
\begin{equation} 
    \ln \left(\frac{K_t}{N}\right) = \frac{S_1L_P}{R} \sinh \left(\frac{t}{R}\right) \gg 1 , 
\end{equation} 
when either of the two indicated inequalities are satisfied.  In the intermediate regime the solution is more complicated.  Of course, since $K_t$ is an integer, these equations can only be satisfied at discrete values of $t$.  Note that $N \geq S_0 \gg 1$, by the boundary conditions on the Milne universe and the assumed validity of the semi-classical approximation to JT dS gravity.  

\section{Conclusions}

We have presented an interpretation of JTdS gravity quite different than any of those previously proposed in the literature.  The static solution of the equations of motion is analogous to the Schwarzschild-de Sitter black hole in higher dimensions: it has two horizons with different entropies\footnote{We've actually argued that the lower entropy should be considered a free energy divided by a temperature, though the precise meaning of this language in the absence of a precisely defined global energy conservation law is not clear.}.  Since the model has only coarse-grained, hydrodynamic degrees of freedom, the decay of the lower entropy state into the higher entropy one is not mediated by Hawking radiation but by a CdL tunneling transition described by the unique Euclidean solution of the equations of motion.  The two Lorentzian continuations of that instanton are a Big Crunch and a cosmology with asymptotically infinity entropy.
The Crunch is interpreted as in \cite{tbcrunch} as the transition of the low entropy state back to the more probable one.  The higher entropy state of the static geometry is found to be the initial condition for a Big Bang cosmology with asymptotically infinite entropy.  We constructed an infinite class of fermionic oscillator models whose coarse-grained dynamics mimics all the relevant features of the classical JT dS picture.  

In the text we have treated this model from the perspective of the causal diamonds along a particular geodesic at some (unspecified) value of $\chi$.  One can view this as the statement that translation in $\chi$ is a gauge symmetry and we have simply chosen a particular gauge.  However, we can also use the same analysis to formulate the model in terms of the Hilbert Bundle formalism of\cite{hilbertbundles}.  If we consider two geodesics at different values $\chi_{1,2}$ and future tips $\tau_{1,2}$ then the intersection between them is the causal diamond of a geodesic and some point between $\chi_1$ and $\chi_2$, whose future tip is at some time $\tau_<$ that is smaller than both $\tau_{1,2}$.  We can assign this a density matrix which is some unitary conjugation of the density matrices of either diamond (since they have the same spectrum, by translation invariance) at this earlier time.  This satisfies the Quantum Principle of Relativity discussed in\cite{hilbertbundles}.   If our coarse-grained descriptions is just JTdS gravity, all of this machinery is just a redundant over-parametrization of the quantum information in the system. However, if we attempt to couple the low energy gravity action to a non-trivial low energy field theory, then the local parametrization of the information will become more useful.  We reserve such investigations for future work.  

The existence of a de Sitter geometry with infinite entropy is a bit surprising and deserves further comment. Although the system has infinite entropy, it is like conventional dS space in the sense that all causal diamonds are related by a symmetry and they are all causally connected.    The model does not have enough degrees of freedom to probe their existence.  However, two dimensional gravity models are often obtained from some sort of dimensional reduction of higher dimensional models, with the $S$ field recording a geometrically defined entropy in the higher dimensions.  In a recent series of papers Collier {\it et. al.} \cite{Collier:2023cyw, Collier:2024kmo, Collier:2024kwt, cemr, Collier:2024mlg, Collier:2025lux, Collier:2025pbm}  have solved the Complex Liouville Theory as a generally covariant theory on Riemann surfaces of genus $g \geq 2$ and proposed an interpretation of the infinite set of correlators they computed as correlators on the boundary at space-like infinity of the geometry
\begin{equation} 
    ds^2 = - dt^2 + \sinh^2 (t/R) dH_{\Gamma , R}^2  , 
\end{equation} 
where $ 0\leq t \leq \infty$ and $dH_{\Gamma , R}^2$ is the metric on the hyperbolic plane of radius $R$, modded out by the Fuchsian group $\Gamma$.  One is also instructed to integrate over the moduli space of Riemann surfaces.  If we cut out a one dimensional boundary on the Riemann surface, the metric on the boundary is $dS_2$.  Although the quantum mechanical interpretation of the correlators in Collier {\it et. al.} has not yet been understood, it seems plausible that they define amplitudes in an infinite dimensional Hilbert space, and that the corresponding boundary Hilbert space might also be infinite dimensional.  We will leave exploration of this idea to future work.

\section{Acknowledgements}

This work was supported in part by the U.S. Dept. of Energy under Grant DE-SC0010008.

\section{Appendix}\label{sec:app}

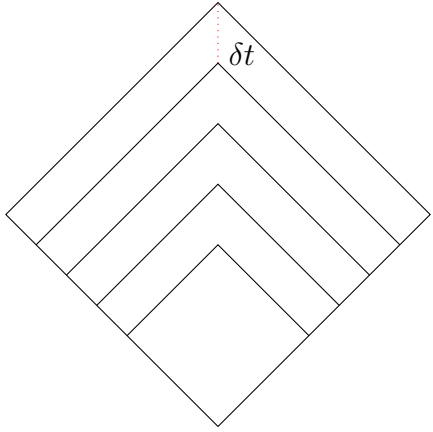
\begin{figure}
    \centering
    \begin{tikzpicture}[scale=0.4]
        \draw[black] (-3,3)--(0,6)--(3,3);
        \draw[black] (-4,4)--(0,8)--(4,4);
        \draw[black] (-5,5)--(0,10)--(5,5);
        \draw[black] (-6,6)--(0,12)--(6,6);
        \draw[black] (0,0)--(-7,7)--(0,14)--(7,7)--(0,0);
        \draw[red,dotted] (0,12)--(0,14);
        \filldraw (0,12.3) circle (0pt) node[anchor=west]{ $\delta t$};
    \end{tikzpicture}
    \caption{A Future Directed Nested Cover of a Diamond. } 
    \label{fig:nested}
\end{figure}

Our fundamental hypothesis about quantum gravity is that the finite entropy predicted by the Bekenstein-Hawking-Jacobson-Fischler-Susskind-Bousso bound for a generic causal diamond implies that the diamond has a finite dimensional Hilbert space. If we consider a nested pair of diamonds with the same past tip and future tips separated by one Planck time, then the two diamonds have Hilbert spaces of different dimension.  For space-time dimension $\geq 3$ and for our JT dS model at $t > R$ the differences in dimension are large.  The mapping of one diamond Hilbert space into the other must be a proper unitary embedding and so the ensemble of maps for a nested sequence of diamonds (Fig. \ref{fig:nested}) is not a one parameter group of unitary maps.  The closest analogy in quantum field theory is the {\it half sided modular inclusion}.  The subalgebra of operators localized in a causal diamond in QFT is Type III$_1$ in the classification of Murray and von Neumann. It has no density matrices because field commutators are singular on the light cone and the diamond is infinitely entangled with its exterior.  Nonetheless, given a choice of vacuum state, each diamond algebra has a modular operator $\Delta = e^{-K}$ with many of the properties of a density matrix.  $\Delta$ is a bounded positive self adjoint operator on the QFT Hilbert space. In particular the {\it modular flow}
\begin{equation}
    a(t) = e^{i K t} a e^{- iKt}  , 
\end{equation}
defines an automorphism of the algebra of the diamond into itself.  For a CFT, $K$ is the integral of a local density times the stress tensor, but in general modular flow is a non-local operation.  However, if we have two diamonds with the same past tip and future tips differing by an infinitesimal amount $\delta t$ as shown in Fig. \ref{fig:nested}, then the inverse of the smaller diamond's flow, followed by the flow of the larger diamond, maps the sub-algebra in the wedge between the two into itself.  This is half sided modular inclusion.  It defines a time dependent evolution on causal diamonds in QFT, an avatar of the quantum gravity evolution discussed in the text.

\end{document}